\documentstyle[11pt,epsfig]{article}
\title{\bf Noncommutative Double Scalar Fields in FRW Cosmology as Cosmical Oscillators}
\author{Behrooz Malekolkalami\footnote{b\_malekolkalami@sbu.ac.ir}\ \ and
         Mehrdad Farhoudi\footnote{m-farhoudi@sbu.ac.ir}\\
        {\small Department of Physics, Shahid Beheshti University, G.C.,}\\
        {\small Evin, Tehran 19839, Iran}}
\begin{document}
\date{\small September 28, 2010}
\maketitle
\begin{abstract}
We investigate the effects caused by noncommutativity of the phase
space generated by two scalar fields that have non--minimal
conformal couplings to the background curvature in the FRW
cosmology. We restrict deformation of the minisuperspace to
noncommutativity between the scalar fields and between their
canonical conjugate momenta. Then, the investigation is carried
out by means of a comparative analysis of the mathematical
properties (supplemented with some diagrams) of the time evolution
of variables in a classical model and the wave function of the
universe in a quantum perspective, both in the commutative and
noncommutative frames. We find that the impose of noncommutativity
causes more ability in tuning time solutions of the scalar fields
and hence, has important implications in the evolution of the
universe. We get that the noncommutative parameter in the momenta
sector is the only responsible parameter for the noncommutative
effects in the spatially flat universes. A distinguishing feature
of the noncommutative solutions of the scalar fields is that they
can be simulated with the well--known three harmonic oscillators
depending on three values of the spatial curvature. Namely the
free, forced and damped harmonic oscillators corresponding to the
flat, closed and open universes, respectively. In this respect, we
call them \textsl{cosmical oscillators}. In particular, in closed
universes, when the noncommutative parameters are small, the
cosmical oscillators have analogous effect with the familiar
beating effect in the sound phenomena. Some of the special
solutions in the classical model and the allowed wave functions in
the quantum model make bounds on the values of the noncommutative
parameters. The existence of a non--zero constant potential (i.e.
a cosmological constant) does~not change time evolutions of the
scalar fields, but modifies the scale factor. An interesting
feature of well--behaved solutions of the wave functions is that
the functional form of the radial part is the same as the
commutative ones within a given replacement of constants caused by
the noncommutative parameters. Furthermore, the Noether theorem
has been employed to explore the effects of noncommutativity on
the underlying symmetries in the commutative frame. Two of the six
Noether symmetries of spatially flat universes, in general, are
retained in the noncommutative case, and one out of the three ones
in non--flat universes.
\end{abstract}
\medskip
{\small \noindent PACS number: $04.50.-h$; $04.20.Fy$;
                               $02.40.Gh$; $04.50.Kd$; $98.80.Qc$; $98.80.Jk$}\newline
 {\small Keywords: Noncommutative Phase Space; Double Scalar Field
                   Cosmology; Quantum Cosmology; Noether Symmetry; Scalar--Tensor Theory.}
\newpage
\bigskip
\section{Introduction}
\indent

Scalar field theories have become generic playgrounds for building
cosmological models related to particle physics~\cite{3}, and more
recently, have played very important contributions in the other
branches of cosmology, e.g., as a powerful tool for providing
expanding accelerated universes. Also, they have key roles in some
models of the early cosmological inflation~\cite{1}, and are
viable favorite candidates for dark matter~\cite{2}. Indeed,
scalar field cosmological models have extensively been studied in
the literature, see, e.g., Refs.~\cite{4} and references therein.

Unifying theories of interactions, such as the Kaluza--Klein,
supergravity and superstring theories, usually predicts
non--minimal couplings between geometry of space--time and scalar
fields. These theories mostly are derived from effective actions
of string theories~\cite{41} or from compactified Kaluza--Klein
theories in four dimensions~\cite{42}. In both cases, resulting
models have non--minimal couplings between gravity and one (or
even more) scalar field. Actually, considering more than one
scalar field has been viewed as an easier approach to study
accelerating models. These cosmological ideas have also been
employed in models of inflation to describe the early
universe~\cite{44} or an evolving scalar field known as
quintessence~\cite{45}. Recently, double scalar--tensor theories
have been applied in extended gravitational theories, and have
given successful descriptions of inflationary
scenarios~\cite{46,47}. Besides, it has been
shown~\cite{qiang2005} that the reduction of a five--dimensional
Brans--Dicke gravity into four dimensions is equivalent to a
4--metric coupled to two scalar fields, which may account for the
present accelerated expansion of the universe. Also, effects of
additional scalar fields plus independent exponential potentials
have been considered in the literature~\cite{48}.

In cosmological models, scalar fields usually present degrees of
freedom and appear as dynamical variables of corresponding
minisuperspaces. This point can be viewed as relevance of
noncommutativity in these models. The proposal of noncommutativity
concept between space--time coordinates, in the first time, was
introduced in 1947~\cite{5}. Thereafter, a mathematical theory,
nowadays known as the noncommutative geometry~(\textbf{NCG}), has
begun to take shape based on this concept since 1980~\cite{6}.
Noncommutativity among space--time coordinates implies
noncommutativity among fields as minisuperspace coordinates. Such
a kind of noncommutativity has gotten more attention in the
literature. In the last decade, study and investigation of
physical theories in the noncommutative~({\bf NC}) frames, such as
the string and M--theory~\cite{7,8}, have caused a renewed
interest on noncommutativity in the classical and quantum
perspectives. In particular, a novel interest has been developed
in studying the NC classical and quantum cosmologies, e.g.
Refs.~\cite{L02K06A08A09} and references therein. Also,
deformation of the phase space structure has been viewed as an
alternative path, in the context of cosmology, toward
understanding quantum gravity~\cite{S03ZFC05KSV10}. The influence
of noncommutativity has been examined by formulation of a version
of the NC cosmology in which a deformation of the minisuperspace
coordinates~\cite{9}--\cite{11} or the minisuperspace
momenta~\cite{12,12a} is required instead of the space--time
coordinates. From a qualitative point of view, noncommutativity in
the minisuperspace coordinates (space sector) leads to general
effects, however, a non--trivial noncommutativity in the momentum
sector introduces distinct and instructive effects in the behavior
of dynamical variables.

In this work, we consider homogeneous and isotropic cosmological
models based on a particular (multi)scalar--tensor gravity theory
with two scalar fields that have non--minimal conformal couplings
to the space--time. The effects of noncommutativity on the phase
space, generated by these fields plus the scale factor, are
investigated. Our approach is to build a NC scenario via a
deformation conveyed by the Moyal product~\cite{7} in a classical
and a quantum perspective, where their cosmological implications
are more appreciated in the classical one. The procedure of
quantization is proceeded by the Wheeler--DeWitt (\textbf{WD})
equation for a wave function of the universe by the Hamiltonian
operator of the model via the generalized Dirac quantization. We
introduce noncommutativity between the scalar fields and between
their canonical conjugate momenta, and will pay more attention to
the outcome of results. Here, we neglect noncommutativity between
scalar fields with the scale factor, such effects can be found in,
e.g., Refs.~\cite{11,12a}. Actually we treat the effects of
noncommutativity by two parameters which are the NC parameters
corresponding to the space and momentum sectors. Then, we analyze
the mathematical properties of solutions and look for relations,
including the ranges and values, among the NC parameters for which
particular or allowed solutions exist.

The manuscript is organized as follows. In Section~$2$, we specify
our toy model and investigate the classical version within the
commutative and NC frames. The quantum version of this model, by
probing the universe wave functions, is considered in Section~$3$,
where the general properties of solutions in the commutative and
NC frames are compared. In Section~$4$, we employ the Noether
theorem and explore the effects of noncommutativity on the
underlying symmetries in the commutative frame. Conclusions are
presented in the last section, and some necessary formulations
have been furnished in the appendix.

\section{The Classical Model}
\indent

We consider a classical model consisting of a cosmological system
presented by a four dimensional action with two non--minimally
scalar fields coupled to gravity in a Friedman--Robertson--Walker
(\textbf{FRW}) universe. To specify the NC effects of the model,
we first treat the commutative version and then consider the NC
one in the following subsections.

\subsection{The Commutative Phase Space}
\indent

A general action for non--minimally coupled double scalar fields
can be written as~\cite{46}
\begin{equation}\label{B1}
{\cal
A}=\int\sqrt{-g}\left\{F(\phi,\psi)R-\frac{1}{2}g^{\mu\nu}\Bigl[A(\phi,\psi)\nabla_\mu\phi
\nabla_\nu\phi+B(\phi,\psi)\nabla_\mu\psi
\nabla_\nu\psi\Bigr]-V(\phi,\psi)\right\}d^4x\hspace{1mm},
\end{equation}
where $g$ is the determinant of the metric $g_{\mu\nu}$, $R$ is
the Ricci scalar, $F(\phi,\psi)$ and $V(\phi,\psi)$ are coupling
and (total) potential functions of scalar fields, respectively.
Also, $A(\phi,\psi)$ and $B(\phi,\psi)$ are two typical functions
coupled to the kinetic terms. We assume homogeneous scalar fields,
that is $\phi=\phi(t)$ and $\psi=\psi(t)$, in the following FRW
metric
\begin{equation}\label{B2}
ds^2=-N^2(t)dt^2+a^2(t)\left(\frac{dr^2}{1-kr^2}+r^2d\Omega^2\right),
\end{equation}
where $N(t)$ is a lapse function, $a(t)$ is a scale factor and $k$
specifies spatial geometry of the universe.

We restrict our considerations to a non--interacting conformally
scalar field model. This {\it ansatz} is obviously restrictive.
However, a general reason for selecting such scalar fields is for
simplicity. Indeed, it allows exact solutions in simple cases, as
those will be discussed in this work and are rich enough to be
useful as a probe for significant modifications that NCG
introduces in the classical and quantum cosmologies. Thus in this
case, one can set $F(\phi,\psi)=f(\phi)+f(\psi)$ where $f$ is a
function of its argument as $f(\chi)=1/(4\kappa)-\xi\chi^2/2$.
Also, $\kappa=8\pi G/c^4$ and $\xi$ is a non--minimal coupling
parameter with an arbitrary value that represents a direct
coupling between the curvature and the scalar fields. The case
$\xi=0$ obviously is the minimally coupling situation; however in
this work, we consider the conformal coupling case, i.e.
$\xi=1/6$, and employ the units $\hbar=1=c$ and $\kappa=3$ (i.e.
$G=3/8\pi$). We also, in general, consider the scalar fields that
are~not subject to any potential\rlap,\footnote{A vanishing
(total) potential, $V(\phi,\psi)$, can also be achieved by
non--interacting (inner) potentials, e.g.
$V(\phi,\psi)=U(\phi)+W(\psi)=\Lambda+(-\Lambda)=0$, where
$\Lambda$ is the cosmological constant. Such a case is an
important one in models with double scalar fields~\cite{46}.}\
 and assume more simple cases with $A(\phi,\psi)=1=B(\phi,\psi)$.
However, at the end of this section, we will investigate the
particular case of non--zero constant values of the potential
function (i.e. the cosmological constant).

Now, based on these assumptions, substituting metric (\ref{B2})
into action (\ref{B1}) yields the Lagrangian density
\begin{equation}\label{B3}
{\cal L}
=\left(kNa-\frac{a\dot{a}^2}{N}\right)\left(1-\frac{\phi^2+\psi^2}{2}\right)+(\phi\dot{\phi}+\psi\dot{\psi})
\frac{a^2\dot{a}}{N}+\frac{a^3}{2N}(\dot{\phi}^2+\dot{\psi}^2)\, ,
\end{equation}
where a total time derivative term has been neglected. By
rescaling the scalar fields as
\begin{equation}\label{rescalingfields}
x=a\phi/\sqrt{2}\,\hspace{1cm} {\rm and} \hspace{1cm}
y=a\psi/\sqrt{2}\, ,
\end{equation}
Lagrangian (\ref{B3}) reads
\begin{equation}\label{B4}
{\cal
L}=kNa-\frac{a\dot{a}^2}{N}+\frac{a(\dot{x}^2+\dot{y}^2)}{N}-\frac{kN(x^2+y^2)}{a}\,
.
\end{equation}
Then, the corresponding Hamiltonian is
\begin{equation}\label{B5,1}
{\cal
H}=N\left[-\frac{p^2_a}{4a}+\frac{p^2_x+p^2_y}{4a}-ka+\frac{k(x^2+y^2)}{a}\right],
\end{equation}
where  $p_a$, $p_x$ and $p_y$ are the corresponding canonical
conjugate momenta. For the conformal time gauge selection, namely
$N=a$, one gets
\begin{equation}\label{B5,2}
{\cal H}=-\frac{p^2_a}{4}+\frac{p^2_x+p^2_y}{4}-ka^2+k(x^2+y^2).
\end{equation}
Therefore, the Hamilton equations are
\begin{eqnarray}\label{B6}
&&\dot{a}=\{a,{\cal H}\}=-\frac{1}{2}p_a\, , \cr
&&\dot{p_a}=\{p_a,{\cal H}\}=2ka\, ,\cr
 &&\dot{x}=\{x,{\cal H}\}=\frac{1}{2}p_x\, , \cr
 && \dot{p_x}=\{p_x,{\cal H}\}=-2kx\, ,\cr
  &&\dot{y}=\{y,{\cal H}\}=\frac{1}{2}p_y\, ,\cr
 &&\dot{p_y}=\{p_y,{\cal H}\}=-2ky\, .
\end{eqnarray}

Solutions of equations (\ref{B6}), with the Hamiltonian constraint
${\cal H} \approx 0$, depend on the curvature index; actually
their solutions are
\begin{eqnarray}\label{C14}
k=1 :\left\{
\begin{array}{lll}
a(t)= A_1\cos t+A_2\sin t,
\hspace{.4cm}x(t)=B_1\cos t+B_2\sin t\hspace{.3cm} \textrm{and} \hspace{4mm}y(t)=C_1\cos t+C_2\sin t\, ,
\\
\\
{\rm with\hspace{.2cm} constraint\!:}\hspace{.7cm}
2(A_1^2+A_2^2)=\sum^{2}_{i=1}(B_i^2+C_i^2)\, ,
\end{array}
\right.
\end{eqnarray}
\begin{eqnarray}\label{C141}
k=-1 :\left\{
\begin{array}{lll}
a(t)= A_3e^t+A_4e^{-t},\hspace{.4cm}x(t)=B_3e^t+B_4e^{-t}\hspace{.3cm} \textrm{and}\hspace{.4cm}y(t)=C_3e^t+C_4e^{-t}\, ,
\\
\\
{\rm with\hspace{.2cm} constraint\!:}\hspace{.7cm} 2A_3A_4=B_3B_4+C_3C_4\,
,
\end{array}
\right.
\end{eqnarray}
\begin{eqnarray}\label{C142}
k=0 :\left\{
\begin{array}{lll}
a(t)=A_5t+A_6,\hspace{.4cm} x(t)=B_5t+B_6\hspace{3mm}\textrm{and}\hspace{4mm}y(t)=C_5t+C_6\, ,
\hspace{.4cm}\,
\\
\\
{\rm with\hspace{.2cm} constraint\!:}\hspace{.7cm} 2A_5^2=B_5^2+C_5^2\, ,
\end{array}
\right.
\end{eqnarray}
where $A_i$'s, $B_i$'s and $C_i$'s are constants of integration.
As it is obvious, if one of the scalar fields vanishes and/or be
equal to each other, i.e. $\phi=\psi(\equiv\chi/\sqrt{2})$, all
equations will lead to those derived in the case of one scalar
field in our previous work~\cite{12a}.

We will compare these solutions with their NC analogs in the next
section.

\subsection{The Noncommutative Phase Space}
\indent

In the classical physics, Noncommutativity is achieved by
replacing ordinary product with the so--called Moyal product,
shown by the $\ast$ notation. This associative  product on
$2l$--dimensional phase space is defined between two arbitrary
functions of general phase space variables, namely
$\zeta^a=(q^i,p^j)$ for $i, j=1, \cdots, l$, as~\cite{7}
\begin{equation}\label{A1}
(f\ast
g)(\zeta)=\exp\left[\frac{1}{2}\alpha^{ab}\partial_a^{(1)}\partial_b^{(2)}\right]f(\zeta_1)g(\zeta_2)
{\biggr|}_{\zeta_1=\zeta_2=\zeta}\, ,
\end{equation}
such that the symplectic structure is
\begin{equation}\label{A2}
(\alpha_{ab})=\left(%
\begin{array}{cc}
\theta_{ij} & \delta_{ij}+\sigma_{ij} \\-\delta_{ij}-\sigma_{ij}& \beta_{ij} \\
\end{array}
\right),
\end{equation}
where $a, b=1, 2, \cdots, 2l$. The matrix elements $\alpha_{ab}$
are assumed to be real, $\theta_{ij}$ and $\beta_{ij}$ are
antisymmetric, and $\sigma_{ij}$ (which can be written as a
combination of $\theta_{ij}$ and $\beta_{ij}$) is symmetric. The
modified (or the $\alpha$--star deformed) Poisson brackets are
defined, in terms of the Moyal product, as
\begin{equation}\label{A3}
\{f,g\}_\alpha=f\ast g-g\ast f\, .
\end{equation}
Hence, the modified Poisson brackets of the phase space variables
are
\begin{equation}\label{A4}
\{q_i,q_j\}_\alpha=\theta_{ij}\,
,\hspace{.5cm}\{q_i,p_j\}_\alpha=\delta_{ij}+\sigma_{ij}
\hspace{.5cm}{\rm and}\hspace{.5cm}\{p_i,p_j\}_\alpha=
\beta_{ij}\, .
\end{equation}

For classical evolution, one starts by proposing that the NC
nature of the minisuperspace variables can be encoded in a
deformation of the Poisson structure. The main advantage of this
approach is that the Hamiltonian does~not need any modification.
Hence, for our model, we require that the algebra of the
minisuperspace variables obeys relations (\ref{A4}) in terms of
the Moyal product defined in (\ref{A1}). However, to introduce
such deformations, it is more convenient to consider a linear and
non--canonical transformation on the classical phase space as (see
e.g. Refs.~\cite{CSjT01,14})
\begin{equation}\label{A42}
x'_i=x_i-\frac{1}{2}\theta_{ij}p^j\hspace{.5cm}{\rm
and}\hspace{.5cm}p'_i=p_i+\frac{1}{2}\beta_{ij}x^j\, .
\end{equation}
Then, if the variables of classical phase space obey the usual
Poisson brackets, i.e. $\{x_i , x_j\} = 0 = \{p_i , p_j\}$ and
$\{x_i , p_j\}=\delta_{ij}$, the primed variables will yield
\begin{equation}\label{A43}
\{x'_i,x'_j\}=\theta_{ij}\,
,\hspace{.5cm}\{x'_i,p'_j\}=\delta_{ij}+\sigma_{ij}
\hspace{.5cm}{\rm and}\hspace{.5cm}\{p'_i,p'_j\}= \beta_{ij}\, ,
\end{equation}
with $\sigma_{ij}=-\theta_{k(i}\beta_{j)l}\delta^{kl}/4$.
Consequently, to consider the noncommutativity effects, one can
work with the ordinary Poisson brackets (\ref{A43}) instead of the
$\alpha$--star deformed Poisson brackets (\ref{A4}). Indeed,
transformation (\ref{A42}) allows an extension of the commutative
classical mechanics to the NC one. In the geometrical language,
the usual Poisson brackets are mapped onto the modified Poisson
brackets through transformation (\ref{A42}) and vice versa. It is
evident that, for a compatible extension, transformation
(\ref{A42}) must be invertible, and this imposes a condition on
the NC parameters which we will specify for our model in below.

Furthermore, let ${\cal H}={\cal H}(x_i,p_i)$ be the Hamiltonian
of a system including the commutative variables; we shift the
canonical variables through (\ref{A42}) and assume that the
functional form of the Hamiltonian in the NC case is still the
same as  the commutative one, i.e.
\begin{equation}\label{C}
{\cal H}_{\rm nc}\equiv{\cal H}(x'_i, p'_i)={\cal
H}\left(x_i-\frac{1}{2}\theta_{ij}p^j,
p_i+\frac{1}{2}\beta_{ij}x^j\right).
\end{equation}
This function is also defined on the commutative space and,
obviously, equations of motion for unprimed variables are
$\dot{x}^i=\partial{\cal H}_{\rm nc}/\partial p_i$\ and
$\dot{p}^i=-\partial{\cal H}_{\rm nc}/\partial x_i$. Evidently,
the effects due to the noncommutativity arise by terms including
the parameters $\theta_{ij}$ and $\beta_{ij}$.

Now, in our model, the following notations are adopted
\begin{center}
$(x^1,x^2,x^3)=(a,x,y)$ \qquad {\rm and} \qquad
$(p^1,p^2,p^3)=(p_a,p_x,p_y)$.
\end{center}
And, in this work, we assume that the only non--zero NC parameters
are\footnote{Note that, noncommutativity between $x$ and $y$ while
they commute with the scale factor is completely consistent with
noncommutativity between the original scalar fields, $\phi$ and
$\psi$, while they also commute with the scale factor.}
\begin{equation}\label{ourassumpyion}
\theta^{23}\equiv\theta\geq0\qquad {\rm and} \qquad
\beta^{23}\equiv4\beta\geq0\, ;
\end{equation}
hence $\sigma^{22}=\sigma^{33}=\theta\beta$, where $\theta$ and
$\beta$ are constants. Actually, in accordance with (\ref{A42}),
we consider the following transformation
\begin{equation}\label{C3}
x\rightarrow
x-\frac{\theta}{2}p_y\hspace{0.1cm},\hspace{0.5cm}p_x\rightarrow
p_x+2\beta y \hspace{0.1cm},\hspace{0.5cm}
y\rightarrow y+\frac{\theta}{2}p_x\hspace{0.4cm} {\rm and}
\hspace{0.4cm}p_y\rightarrow p_y-2\beta x.
\end{equation}
Transformation (\ref{C3}) dictates that it is inverted when its
determinant is~not zero, that is when $\theta\beta\neq 1$. Then,
by making transformation (\ref{C3}) in Hamiltonian (\ref{B5,2}),
the NC Hamiltonian is
\begin{equation}\label{C4}
{\cal H}_{\rm
nc}=-\frac{p^2_a}{4}+\frac{1+k\theta^2}{4}(p_x^2+p_y^2)-ka^2+(k+\beta^2)(x^2+y^2)+(k\theta+\beta)(y
p_x-x p_y)\, ,
\end{equation}
and equations of motion are
\begin{equation}\label{C5}
\dot{a}=\{a,{\cal H}_{\rm nc}\}=-\frac{1}{2}p_a\,
,\hspace{1cm}\dot{p_a}=\{p_a,{\cal H}_{\rm nc}\}=2ka\, ,
\end{equation}
and
\begin{eqnarray}\label{C9}
 &&\dot{x}=\{x,{\cal H}_{\rm nc}\}=\frac{1}{2}(1+k\theta^2)p_x+(k\theta+\beta)y\, ,\cr
 &&\dot{p_x}=\{p_x,{\cal H}_{\rm nc}\}=-2(k+\beta^2)x+(k\theta+\beta)p_y\, , \cr
 &&\dot{y}=\{y,{\cal
 H}_{\rm nc}\}=\frac{1}{2}(1+k\theta^2)p_y-(k\theta+\beta)x
   \, , \cr
 &&\dot{p_y}=\{p_y,{\cal H}_{\rm nc}\}=-2(k+\beta^2)y-(k\theta+\beta)p_x\, .
\end{eqnarray}
Equations (\ref{C9}) show that, in general, the motion equations
of the scalar fields are coupled in the NC case, and these
equations reduce to the commutative equations (\ref{B6}) for
$\theta=0=\beta$, as expected. Also, in the purposed model, the
noncommutativity does~not affect the time dependence of the scale
factor and its solution is the same as the commutative case.

The motion equations of the scalar fields, after eliminating
momenta variables in (\ref{C9}), are
\begin{equation}\label{C13}
\ddot{x}=-k(1-\theta\beta)^2x+2(k\theta+\beta)\,\dot{y}
\hspace{1cm}{\rm and}
\hspace{1cm}\ddot{y}=-k(1-\theta\beta)^2y-2(k\theta+\beta
)\,\dot{x}\hspace{.5mm},
\end{equation}
with the Hamiltonian constraint
\begin{center}
\hspace{1.45cm}$x^2+y^2={\rm constant}$\hspace{1.5cm}{\rm if}
\hspace{.2cm} $1+k\theta^2=0$\footnote{The condition
$1+k\theta^2=0$ is possible only when $k=-1$, and hence
$\theta=1$.}
\end{center}
or
\begin{center}
$(\dot{x}^2+\dot{y}^2)+k(1-\theta\beta)^2(x^2+y^2)={\rm
constant}$\hspace{1.5cm}{\rm if}
\hspace{.3cm}$1+k\theta^2\neq0$.{}\hspace{2.75cm}{}\qquad{}
\end{center}
Space--time geometries with $k=0, 1$ only satisfy the latter
constraint for any $\theta$, while $k=-1$ geometry fulfills the
former constraint with $\theta=1$ (and consequently, because of
the inversion condition, $\beta\neq 1$), and the latter one with
$\theta\neq 1$. In general, solutions of (\ref{C13}) depend on the
sign of a quantity defined as
$\Delta\equiv(1+k\theta^2)(\beta^2+k)$.

For $k=0,1$, the sign of $\Delta$ is always positive, and thus,
the real solution of equations (\ref{C13}) can be written as
\begin{equation}\label{C16}
k=0, 1 :\!\cases{x(t)= A\cos\omega_1t+B\cos\omega_2t \cr
 y(t)=A\sin\omega_1t+B\sin\omega_2t\, ,\cr}
\end{equation}
where $A$ and $B$ are constants of integration subject to the
corresponding Hamiltonian constraint, and
\begin{equation}\label{C161}
\omega_1\equiv \sqrt{\Delta}-k\theta-\beta\qquad\quad {\rm
and}\qquad\quad \omega_2\equiv-\sqrt{\Delta}-k\theta-\beta.
\end{equation}
Obviously, the corresponding terms in (\ref{C16}) are in $\pi/2$
phase difference.

In the case of $k=0$, the NC parameter $\theta$ does~not actually
appear in equations (\ref{C13}), and hence, in a spatially flat
FRW universe, the scalar fields motion equations are affected only
by the NC parameter $\beta$. Besides, one gets $\omega_1=0$ and
$\omega_2=-2\beta$, and thus solutions are similar to the motion
of a free harmonic oscillator. However, the time dependence of
solutions are modified from linear in the commutative case, to
periodic in the NC case with the period of $\pi/\beta$ which
allows the model to be adjusted more easily with the observational
data.

For $k=1$ geometry, the commutative solutions (\ref{C14}) are
simple harmonics with the period of $2\pi$, whereas, the NC
solutions (\ref{C16}) in general are~not, though they still
oscillate between two limits. Incidentally, solutions (\ref{C16})
are symmetric with respect to the NC parameters, and results
do~not vary by replacement of their roles. Besides, as
$\omega_1\neq\omega_2$, each solution of (\ref{C16}) can be
considered as a forced (driven) harmonic oscillator, such that if
the ratio of $\omega_2/\omega_1$ is a rational fraction, then
solutions will be periodic (as in the Lissajous figures) with
angular frequency given by the greatest common divisor of
$\omega_1$ and $\omega_2$. Otherwise, solutions are non--periodic
and never repeat themselves. A general condition that picks
periodic solutions has been obtained in the appendix (supplemented
with a diagram). Meanwhile, in the following special example, we
provide a better insight about this situation.

Without loss of generality, let us consider the special case
$\beta=0$, with $A=-B\equiv D/2$, for which solutions (\ref{C16})
read
\begin{equation}\label{C17}
\!\cases{x(t)= D\sin\left(\theta
t\right)\sin\left(\sqrt{1+\theta^2}\, t\right) \cr
y(t)=D\cos\left(\theta t\right)\sin\left(\sqrt{1+\theta^2}\,
t\right), \cr}
\end{equation}
with initial conditions $x(0)=0=y(0)$,\ $\dot{x}(0)=0$ and
$\dot{y}(0)=D\sqrt{1+\theta^2}$. The condition for the periodic
solutions is
\begin{center}
$\omega_2/\omega_1=-\frac{\sqrt{1+\theta^2}+\theta}
{\sqrt{1+\theta^2}-\theta}=-n_2/n_1$ ,
\end{center}
where $n_1$ and $n_2$ are positive integers and obviously
$n_2>n_1$. By solving this equation in terms of $\theta$, one gets
\begin{center}
$\theta=\frac{1}{2}\left(n_2-n_1\right)/\sqrt{n_1n_2}=\frac{1}{2}
\left(\sqrt{n_2/n_1}-\sqrt{n_1/n_2}\/\right)$.
\end{center}
This relation provides different values of $\theta$ parameter in
terms of two integers, for which solutions (\ref{C17}) can be
periodic. It is constructive to plot $\theta$ in terms of $r\equiv
n_2/n_1$ as a continuous quantity\rlap,\footnote{Note that,
$n_2/n_1$ is a rational number greater than unity, indeed, $r=1$
yields $\theta =0$ which resumes the commutative case.}\
 and this has been illustrated in
Fig.~$1$ for range $1\leq r<80$. As it is obvious, $d\theta/dr$ is
always positive, thus the function $\theta(r)$ changes
monotonically with $r$. Also, when $r$ has very large values, the
rate $d\theta/dr$ approximately decreases as $1/\sqrt{r}$. For the
smallest allowed values of integers, i.e. $n_2=2$ and $n_1=1$, we
get the smallest value of $\theta$ parameter, i.e. $\theta_{\rm
min}=1/2\sqrt{2}$ for the periodic solutions. In solutions
(\ref{C17}), the terms $D\sin(\theta t)$ and $D\cos(\theta t)$ are
envelopes of the corresponding curves. These types of
oscillations, due to the periodic variation of the amplitude, are
usually called beats. The well--known example of such oscillations
is the simple harmonic one, where a driven force causes mechanical
beats. Also, this situation can be simulated in the acoustical
systems that produce a sound effect, known as beating~\cite{71},
when $|\omega_2|-|\omega_1|=2\theta$ is sufficiently small and the
terms including $\theta^2$ in (\ref{C17}) are ignored. As a sample
illustration of the beating effect for these {\it cosmical
oscillators}, we have graphed a plot of $x(t)$, equations
(\ref{C17}), with its envelopes $\pm D\sin(\theta t)$, in Fig.~$2$
(right) for numerical values $\theta=0.1$ and $D=1$. As the figure
illustrates, beats are more obvious when the envelopes are drawn.
%%%%%%%%%%%%%%%%%%%%%%%%%%%%%%%%%%%%%%%%%%%%%%%%%%%%1111
\begin{figure}
\begin{center}
\epsfig{figure=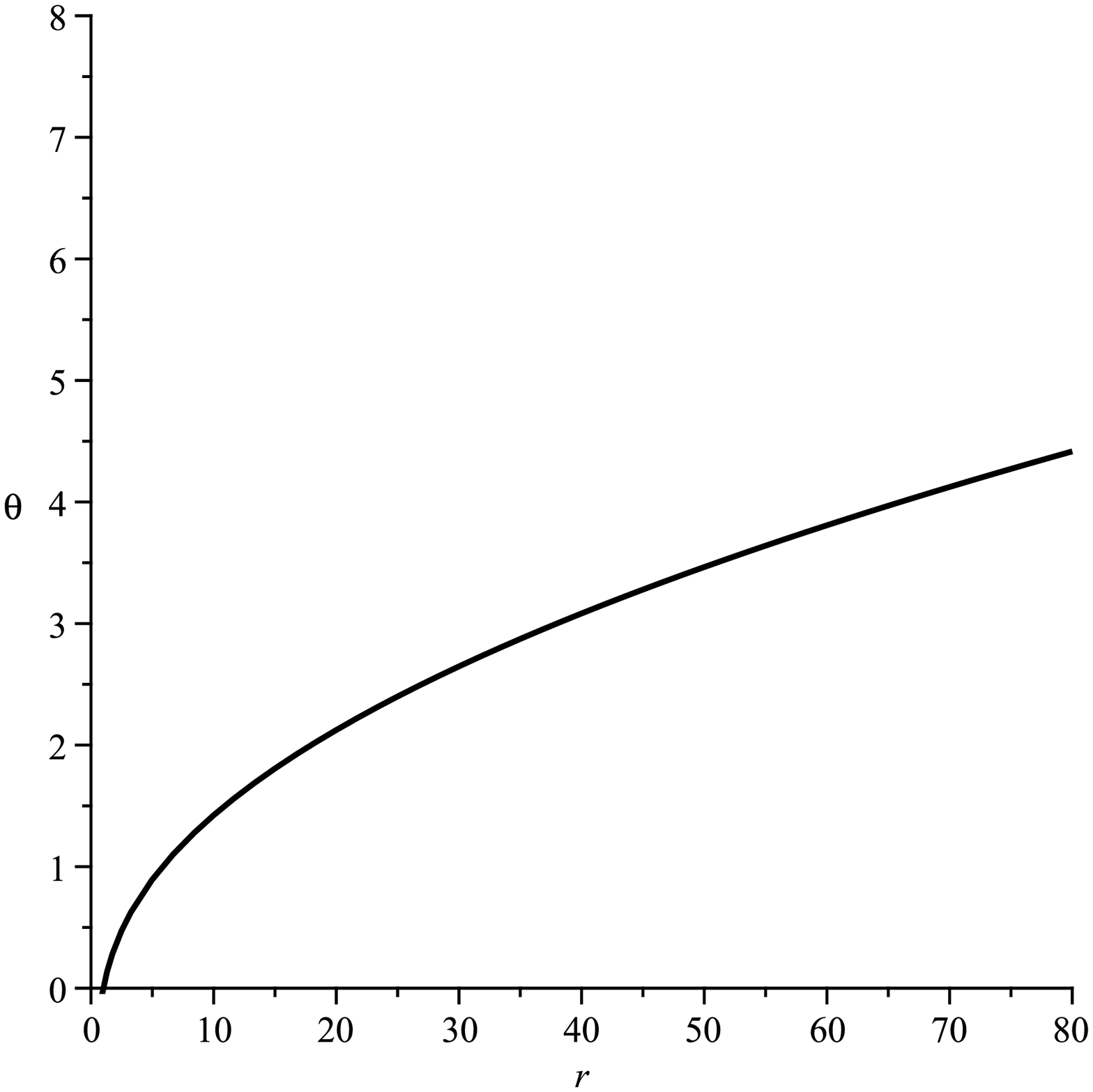,width=7cm}

{\footnotesize \textbf{Fig. 1}: The NC parameter $\theta$, in
$k=1$ case with $\beta =0$, as a function of $r\equiv n_2/n_1$.
The points whose $r$ are rational

fractions correspond to the periodic scalar fields, and other
points correspond to non--periodic ones.}
\end{center}
\end{figure}
%%%%%%%%%%%%%%%%%%%%%%%%%%%%%%%%%%%%%%%%%%%%%%%%%%%%%%%%%2222
\begin{figure}
\includegraphics[scale=0.3]{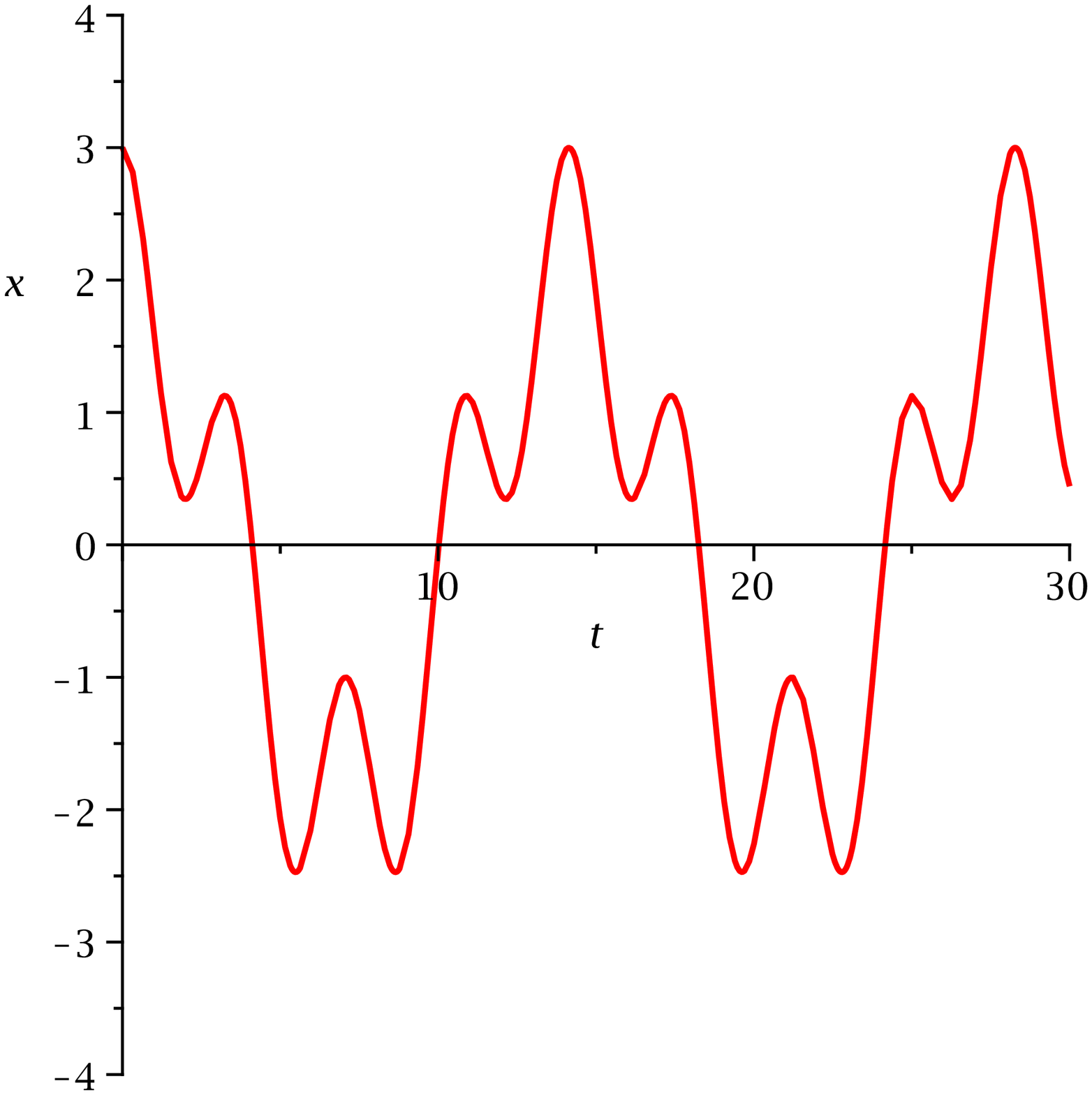}\hspace{3.2cm}
\includegraphics[scale=0.3]{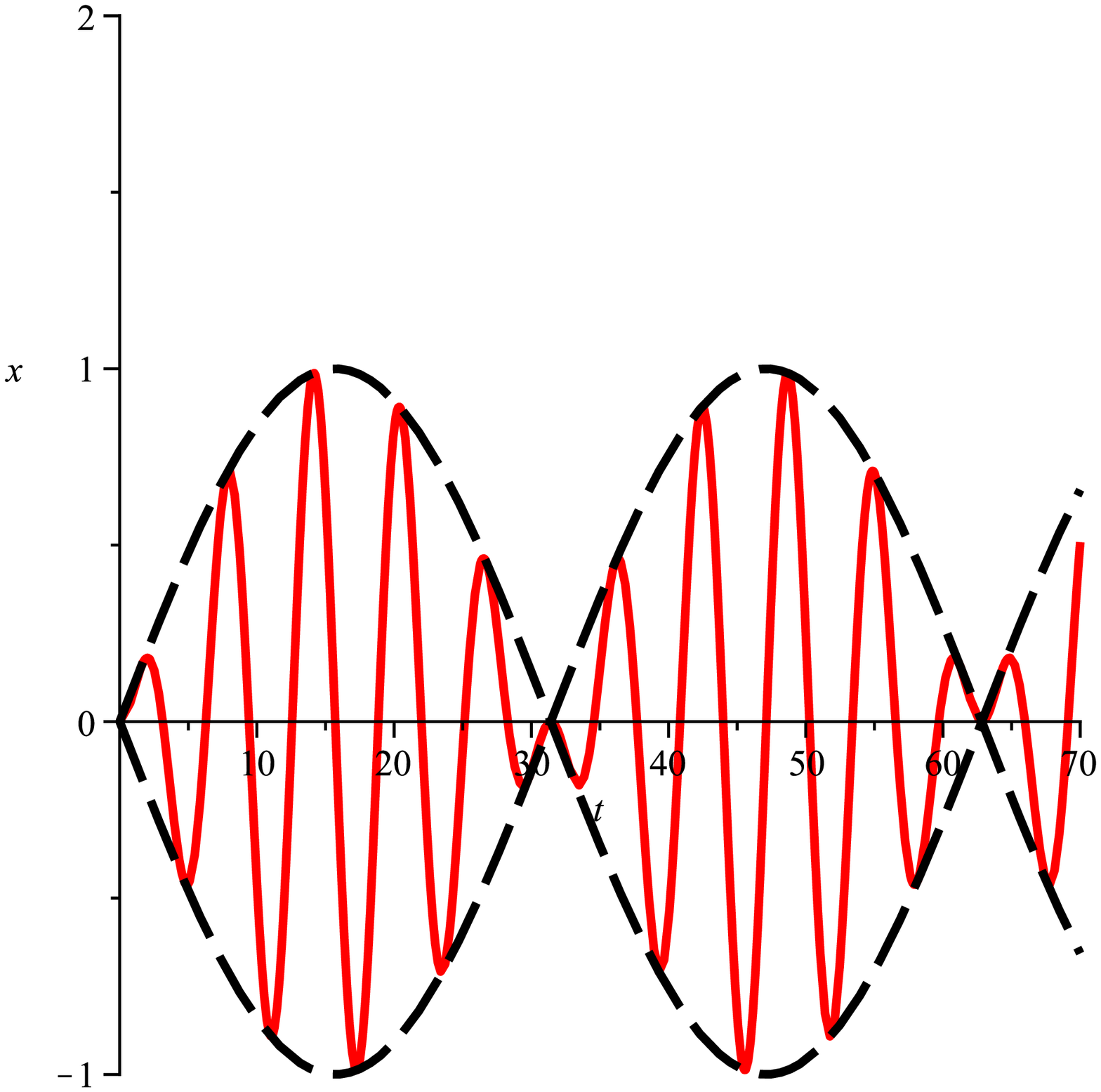}
\begin{center}
{\footnotesize \textbf{Fig. 2}: The NC case for $k=1$: periodic
field (left) with $\theta=2=\beta $, beating effect (right) with
$\theta=0.1$ and $\beta=0$

including the cosmical oscillator (solid line) and envelops
(dashed lines).}
\end{center}
\end{figure}
%%%%%%%%%%%%%%%%%%%%%%%%%%%%%%%%%%%%%%%%%%%%%%%%%%%%%%%%%%%%%

When the ratio of $\omega_2$ and $\omega_1$ is~not a rational
fraction, solutions are non--periodic but, as periodic cases,
their behaviors still depend on the values of the NC parameters.
For example, if one constructs a plot with numerous or a few
relative extremum in a given time interval, then iterative
drawings will indicate that the separation between the high and
low points increases when the NC parameters tend to smaller
values. This property is intensified for values less than unity,
which is illustrated in Fig.~$3$ for $y(t)$ in equations
(\ref{C16}) with constants $A=2$ and $B=1$. From this point of
view, the NC solutions have particular preference with respect to
the corresponding commutative ones.
%%%%%%%%%%%%%%%%%%%%%%%%%%%%%%%%%%%%%%%%%%%%%%%%%%%%%%%%%33333
\begin{figure}
\begin{center}
\epsfig{figure=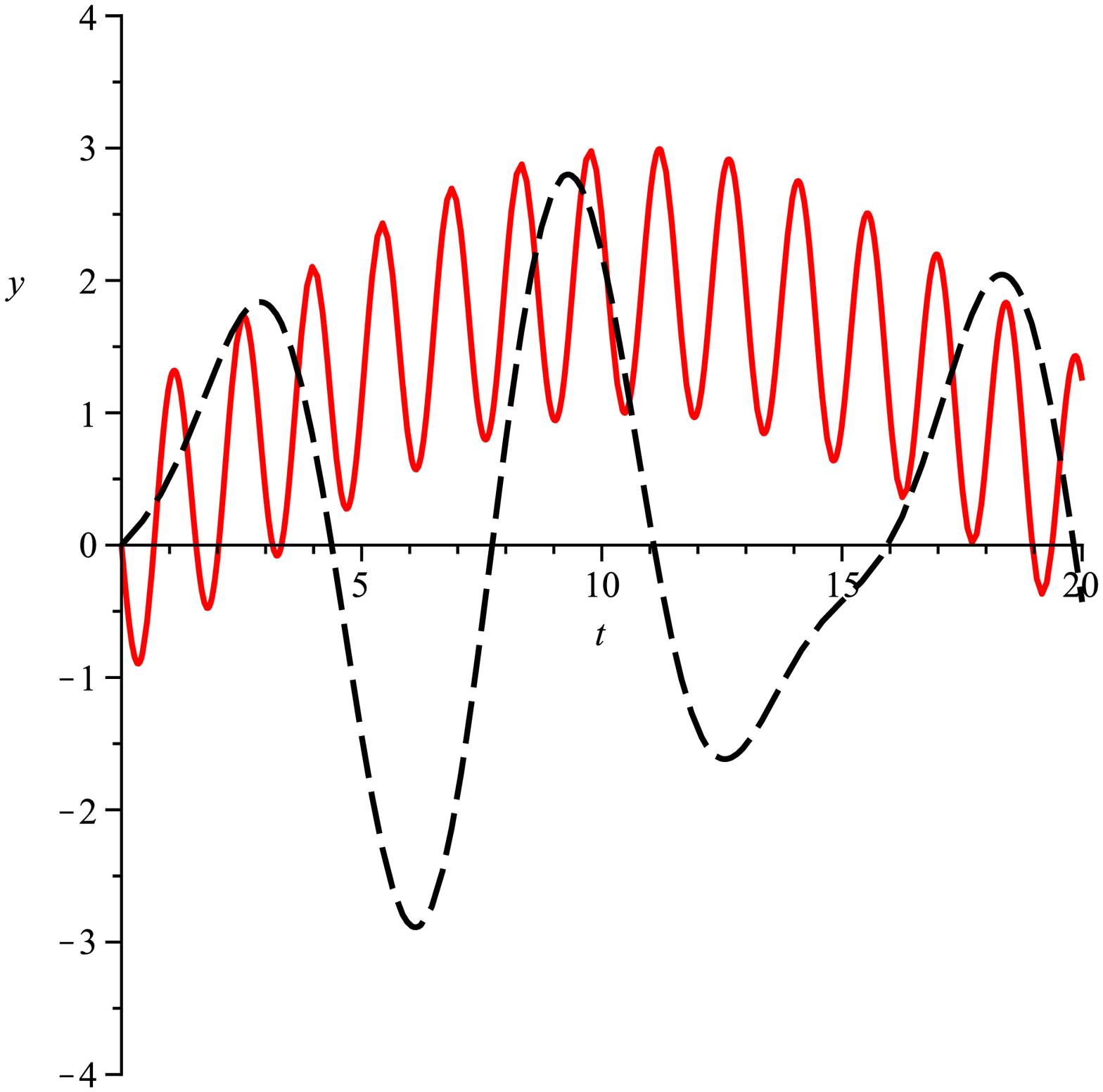,width=7cm}

{\footnotesize \textbf{Fig. 3}: Non--periodic field in the NC case
for $k=1$: with $\theta=2$ and $\beta=0.1$ (solid line),

and with $\theta=0.2$ and $\beta=0.01$ (dashed line).}
\end{center}
\end{figure}
%%%%%%%%%%%%%%%%%%%%%%%%%%%%%%%%%%%%%%%%%%%%%%%%%%%%%%%%%%%%%%

For $k=-1$ geometries, one has $\Delta=(1-\theta^2)(\beta^2-1)$
that can be either positive, or negative or zero depending on
different choices of the NC parameters. That is, when ($\theta>1$
and $\hspace{2mm}\beta<1$) or ($\theta<1$ and
$\hspace{2mm}\beta>1$), $\Delta$ is positive and hence one gets
solutions of type (\ref{C16}). However, in this case, the NC
parameters have upper and lower bounds, and there is more
restriction on finding the periodic solutions than in $k=1$
geometry. On the other hand, when $\Delta$ is negative and
$\theta\neq\beta$, the real solutions of equations (\ref{C13}) are
\begin{equation}\label{C15}
k=-1:\!\cases{
x(t)=\left(C\sinh\sqrt{-\Delta}t+
D\cosh\sqrt{-\Delta}t\right)\cos(\theta-\beta)t
\cr
y(t)=\left(C\sinh\sqrt{-\Delta}t+
D\cosh\sqrt{-\Delta}t\right)\sin(\theta-\beta)t,
\cr}
\end{equation}
where $C$ and $D$ are constants of integration subject to the
corresponding Hamiltonian constraint. Solutions (\ref{C15}) are
constrained to the condition that both NC parameters
simultaneously have a lower bound, namely $\theta$  and $\beta>1$,
or an upper bound, $\theta$ and $\beta<1$. Also, in the case
$k=1$, oscillations of the scalar fields in (\ref{C15}) have the
phase difference of $\pi/2$. These solutions, in contrast to their
commutative analogs, are oscillating with a time hyperbolic
amplitude that depends on the NC parameters and gives enough room
for better adjustments. For instance, one can change the time
interval between two successive zero points of the scalar fields,
for the interval is $\pi/|\theta-\beta|$. Besides, increasing and
decreasing amplitudes in (\ref{C15}) with respect to the time
depend on initial conditions. As an example, if one chooses the
initial condition $C=-D$, then solutions of (\ref{C15}) will
oscillate with decreasing amplitudes. Such solutions are similar
to the damped harmonic oscillators with amplitudes proportional to
$\exp(-\sqrt{-\Delta}t)$ as envelopes. The main characters of such
an oscillator, namely the {\it time decay} and the {\it natural
frequency}, can be described in terms of the NC parameters as
$\tau=1/\sqrt{-\Delta}$ and
$\omega_0=\sqrt{\omega_1\omega_2}=|1-\theta\beta|$, respectively.
Another interesting point is that, if one assumes $\theta$ and
$\beta >1$ then, solutions will damp quickly. Inversely, to
possess more lately damped oscillations, the best choice of the NC
parameters in the range $\theta$ and $\beta<1$ is when one assumes
$\theta\rightarrow1^-$  and $\beta\rightarrow0$, in which each of
the scalar fields has a maximum number of oscillations before the
complete damping occurs. The diagram of such an oscillation is
plotted in Fig.~$4$ for the $x(t)$ of solutions (\ref{C15}) with
$D=1$ for numerical values ($\theta=1.5$ and
$\hspace{1mm}\beta=1.05 $) as quick damping, and for
($\theta=0.999$ and $\hspace{1mm}\beta=0.001$) as late damping.

When $\theta=\beta$ in $k=-1$ geometry, solutions of (\ref{C13})
are decoupled hyperbolic time functions with the coefficient
$(1-\theta^2)$ in the exponent, which is the only difference
between the NC and commutative solutions. The case $\Delta=0$ is
also possible when $\theta=1$ or when $\beta=1$, where solutions
(\ref{C15}) are again periodic with the period of $2\pi/|\beta-1|$
or $2\pi/|\theta-1|$, respectively. Note that, the case $k=-1$
with $\theta=1$ resembles the constraint condition $1+k\theta^2=0$
with arbitrary $\beta$, and however for $\beta=1$ one gets trivial
constant solutions, but this value is~not allowed.
%%%%%%%%%%%%%%%%%%%%%%%%%%%%%%%%%%%%%%%%%%%%%%%%%%%%%%%%%%%44444
\begin{figure}
\begin{center}
\epsfig{figure=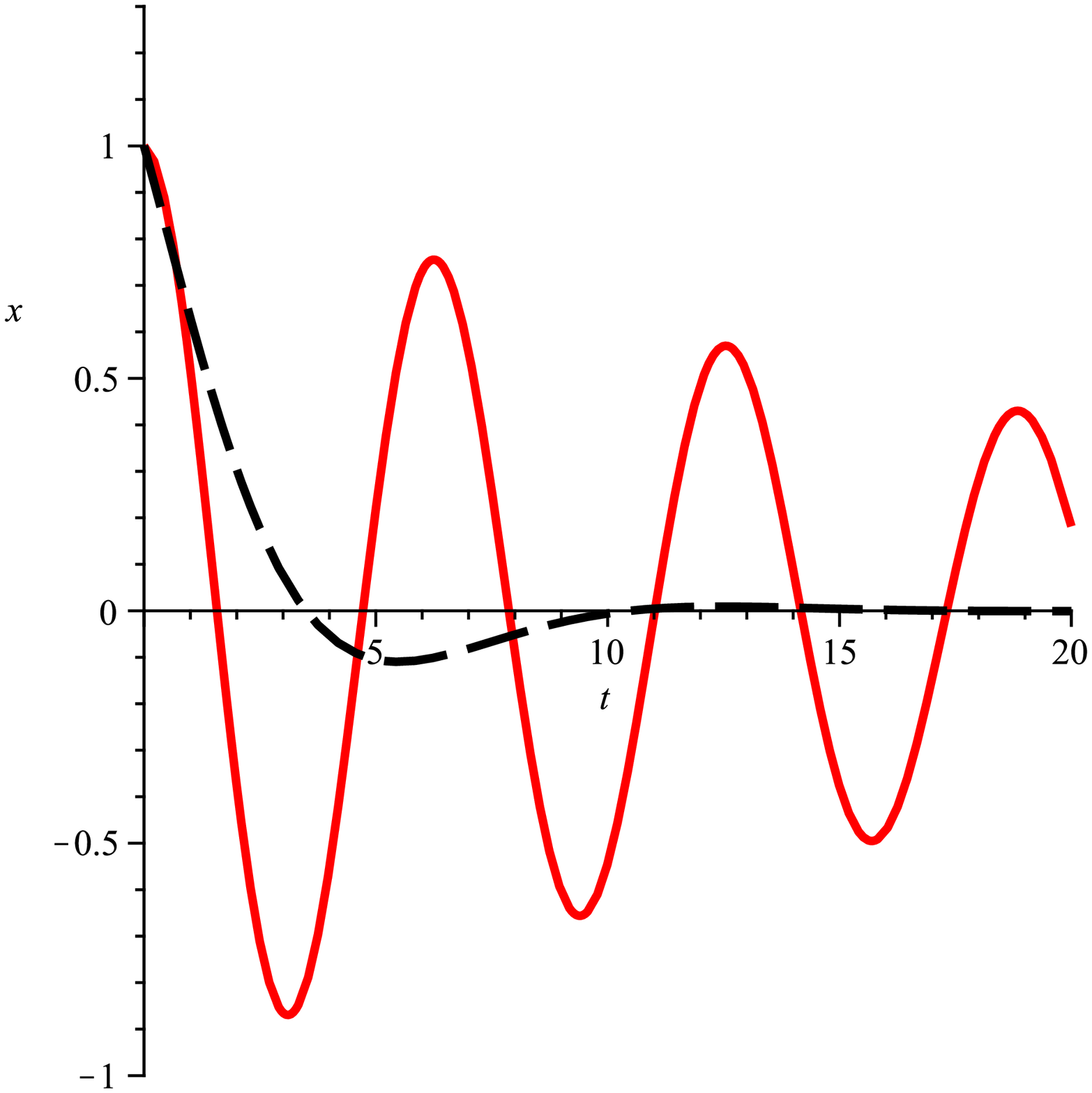,width=7cm}

{\footnotesize \textbf{Fig. 4}: Damping scalar field in the NC
case $k=-1$: late damping with $\theta=0.999$ and $\beta=0.001$
(solid line), and

quick damping with $\theta=1.5$ and $\beta=1.05$ (dashed line).}
\end{center}
\end{figure}
%%%%%%%%%%%%%%%%%%%%%%%%%%%%%%%%%%%%%%%%%%%%%%%%%%%%%%%%%%%%%%%%%%%%%%

\medskip
\noindent{\bf Non--Zero Constant Potential}
\medskip

Let us obtain the equations of motion for the scale factor when
one considers a non--zero constant value for the potential
function (i.e. a cosmological constant) in action (\ref{B1}).
Thus, suppose the potential function is $V(\phi,\psi)=\Lambda$. In
a manner similar to the free--potential case, one gets the
Hamiltonian
\begin{equation}\label{B5,111}
{\cal
H}=-\frac{p^2_a}{4}+\frac{p^2_x+p^2_y}{4}-ka^2+k(x^2+y^2)+\Lambda
a^4\, ,
\end{equation}
in the commutative case, and
\begin{equation}\label{C41}
{\cal H}_{\rm
nc}=-\frac{p^2_a}{4}+\frac{1+k\theta^2}{4}(p_x^2+p_y^2)-ka^2+(k+\beta^2)(x^2+y^2)+(k\theta+\beta)(y
p_x-x p_y)+\Lambda a^4\, ,
\end{equation}
in the NC one. It is obvious that the Hamiltonian equations for
the scalar fields, derived from (\ref{B5,111}) and (\ref{C41}),
are the same as those obtained in the corresponding
free--potential case, namely equations (\ref{B6}) and (\ref{C9}),
respectively. But, the Hamiltonian equations of the scale factor,
for both the commutative and NC cases, are modified in the same
manner as
\begin{equation}
\dot{a}=-\frac{p_a}{2}\hspace{1.2cm}{\rm
and}\hspace{1.2cm}\dot{p_a}=2ka-4\Lambda a^3\, .
\end{equation}
Eliminating the momentum variable gives
\begin{eqnarray}
&&\ddot{a}=2\Lambda a^3-k a.
\end{eqnarray}
By integrating, one gets
\begin{eqnarray}\label{CD}
&&\dot{a}^2=\Lambda a^4-ka^2+ v_0^2\, ,
\end{eqnarray}
where $v_0^2$ is an integration constant. Solution of equation
(\ref{CD}) can be written in terms of the Jacobi elliptic
functions, that is
\begin{equation}\label{CD1}
a(t)=\frac{v_0}{\alpha}\hspace{.6mm}{\bf\textsf{sn}}
\left(\alpha\hspace{.6mm}t,m\right),
\end{equation}
where the initial condition $a(0)=0$,
\begin{equation}\label{CD2}
\alpha^2\equiv\frac{1}{2}\left(k+\sqrt {{k}^{2}-4\,\Lambda
v_0^2}\right)\hspace{10mm}{\rm and}\hspace{10mm}
m\equiv\left[(\alpha^2/\Lambda v_0^2)-1\right]^{-1/2}.
\end{equation}
The Jacobi function, on the right hand side (\ref{CD1}), behaves
as a sine function when $\alpha$ is a real and $m$ is either a
real or a pure imaginary number\rlap.\footnote{For properties of
the elliptic functions, see, e.g., Refs.~\cite{16}.}\
 Therefore, for negative potentials in an arbitrary geometry, and also
for positive potentials in the $k=1$ geometry when $4\,\Lambda
v_0^2\leq 1$, the scale factor behaves periodically as a
sinusoidal function. On the other hand, for positive potentials in
$k=0, -1$ geometries, and in the $k=1$ geometry when $4\,\Lambda
v_0^2> 1$, the Jacobi function has a complex value, and hence
these situations are~not allowed.

In the limit $\Lambda\rightarrow 0$ for $k=1$, one gets
$m\rightarrow 0$, $\alpha^2\rightarrow1$ and ${\bf\textsf{sn}}(t,
0)$ behaves as $\sin(t)$. Hence, we get back to the
free--potential solution (\ref{C14}), as expected. However, in
such a limit, for $k=-1, 0$, as $\alpha\rightarrow 0$, solution
(\ref{CD1}) is~not valid, and hence, one must consider equation
(\ref{CD}). Now, by regarding the limit $\Lambda\rightarrow 0$ in
equation (\ref{CD}) for $k=-1, 0$, and solving the resulted
equation, one again gets the free--potential solutions
(\ref{C141}) and (\ref{C142}), respectively.

\section{The Quantum Model}
\indent

Although the effects of NC cosmology are mostly desired and
important in classical approaches, but still it is instructive to
investigate its quantum counterparts. In particular, if a universe
has been commenced by a big bang or from a very very tiny scale,
then it would be plausible that quantum behaviors should have
significant implications in its evolution and cosmology. Thus, in
this section we proceed to quantize the cosmological model given
by action (\ref{B1}) for free--potential, such that the canonical
quantization of the phase space leads to the WD equation,
$\hat{{\cal H}}\Psi=0$, where $\hat{{\cal H}}$ is the Hamiltonian
operator and $\Psi$ is a wave function of the universe~\cite{13}.
We employ the usual canonical transition from classical to quantum
mechanics via the generalized Dirac quantization of the Poisson
brackets to quantum commutators, i.e.
$\{\}\rightarrow-i[\hspace{.1cm}]$. Then, as the classical
approach, we investigate the general properties of the wave
function in the commutative and NC frames of the quantum model in
the following subsections.

\subsection{The Commutative Quantum Cosmology}
\indent

As usual, the operator form of Hamiltonian (\ref{B5,2}) can be
acquired by the replacements $p_a\rightarrow -i\partial_a$,
$p_x\rightarrow -i\partial_x$ and $p_y\rightarrow -i\partial_y$.
Assuming a particular factor ordering, the corresponding WD
equation is
\begin{equation}\label{D1}
\left[\frac{\partial^2}{\partial
a^2}-\left(\frac{\partial^2}{\partial
x^2}+\frac{\partial^2}{\partial
y^2}\right)+4k\left(x^2+y^2-a^2\right)\right]\Psi(a,x,y)=0\, .
\end{equation}
In terms of the polar coordinates
\begin{equation}\label{D2}
x=\rho\cos\varphi\hspace{1cm}{\rm and}\hspace{1cm}y=\rho\sin\varphi,
\end{equation}
equation (\ref{D1}) reads
\begin{equation}\label{D3}
\left[\partial^2_a-4ka^2-\left(\partial^2_\rho+\frac{\partial_\rho}{\rho}+
\frac{\partial_\varphi^2}{\rho^2}\right)+4k\rho^2\right]\Psi(a,\rho,\varphi)=0.
\end{equation}
Let us choose a product {\it ansatz} as a solution of equation
(\ref{D3}), namely
\begin{equation}\label{D4}
\Psi(a,\rho,\varphi)=A(a)B(\rho)e^{2i \nu\varphi},
\end{equation}
where $\nu$ is a real constant. By substituting (\ref{D4}) into
equation (\ref{D3}), one gets
\begin{equation}\label{D41}
A''+4\left(\mu-ka^2\right)A=0
\end{equation}
and
\begin{equation}\label{D42}
\rho^2B''+\rho B'+4\left(\mu\rho^2-k\rho^4-\nu^2\right)B=0,
\end{equation}
where the prime denotes ordinary derivative with respect to the
argument and $\mu$ is a constant of separation. Solutions of
equations (\ref{D41}) and (\ref{D42}) for real values of $\mu$,
corresponding to the curvature index and boundary conditions, are
\begin{eqnarray}\label{C1411}
k=0 :\left\{
\begin{array}{lll}
A_\mu \left( a \right) =C_1\sin \left(2\sqrt{\mu}\, a\right)
+C_2\cos \left(2\sqrt{\mu}\, a\right)
\\
\\
B_{\mu\nu} \left( \rho \right)
=D_1J_{2\nu}\left(2\sqrt{\mu}\,\rho\right)+D_2Y_{2\nu}\left(2\sqrt{\mu}\,\rho\right)
\end{array}
\right.
\end{eqnarray}
and
\begin{eqnarray}\label{C14111}
k=-1, 1 :\left\{
\begin{array}{lll}
A_\mu\left( a \right) =a^{-1/2}\left[C_3 M_{{\frac {\mu}{2\sqrt
{k}}}, \frac{1}{4}}\left(2\sqrt{k}\,{a}^{2}\right) +C_4 W_{{\frac
{\mu}{2\sqrt {k}}},
\frac{1}{4}}\left(2\sqrt{k}\,{a}^{2}\right)\right] \,
\\
\\
B_{\mu\nu} \left( \rho \right)=\rho^{-1}\left[D_3 M_{{\frac
{\mu}{2\sqrt {k}}},\nu}\left(2\sqrt{k}\,{\rho}^{2}\right) +D_4
W_{{\frac {\mu}{2\sqrt {k}}},
\nu}\left(2\sqrt{k}\,{\rho}^{2}\right)\right] ,
\end{array}
\right.
\end{eqnarray}
where $J_{2\nu}$ and $Y_{2\nu}$ are respectively the Bessel
functions of the first and second kind, $M_{\eta,\lambda}$ and
$W_{\eta,\lambda}$ are the Whittaker functions and, $C_i$'s and
$D_i$'s are constants. For positive (zero) curvature, the
Whittaker (Bessel) term $M_{\eta,\lambda}$ $(Y_{2\nu})$, in the
classically forbidden region, is divergent\rlap.\footnote{For the
properties of the Whittaker and Bessel functions and their
indices, see again Refs.~\cite{16}.}\
 Thus, one can discard these terms and write the well--defined
eigenfunctions of equation (\ref{D3}) as
\begin{eqnarray}\label{D71}
\psi_{\mu\nu}(a,\rho)=A_\mu(a)B_{\mu\nu}(\rho)\propto\left\{\begin{array}{lll}
\Big[C_1\sin \left(2\sqrt{\mu}\, a\right) +C_2\cos
\left(2\sqrt{\mu}\,
a\right)\Big]J_{2\nu}\left(2\sqrt{\mu}\,\rho\right)\hspace{1.2cm}{\rm
for}~~~~k=0,
\\
\\
(\rho\sqrt{a})^{-1}W_{\frac{\mu}{2},
\frac{1}{4}}\left(2{a}^{2}\right)W_{\frac{\mu}{2},
\nu}\left(2{\rho}^{2}\right)\hspace{3.65cm}{\rm for}~~~~k=1,
\\
\\
(\rho\sqrt{a})^{-1}M_{-i\frac{\mu}{2},
\frac{1}{4}}\left(2i{a}^{2}\right)M_{-i\frac{\mu}{2},
\nu}\left(2i{\rho}^{2}\right)\hspace{2.7cm}{\rm for}~~~~k=-1,
\end{array}
\right.
\end{eqnarray}
where in $k=-1$, for simplicity, we have written solution
(\ref{D71}) only in terms of the Whittaker function
$M_{\eta,\lambda}$. The wave packet corresponding to (\ref{D71})
is
\begin{equation}\label{D72}
\Psi(a,\rho,\varphi)=\int^\infty_{-\infty}\int^\infty_{-\infty} E_{\mu}E_{\nu}
\psi_{\mu\nu}e^{2i\nu\varphi}d\mu d\nu\,,
\end{equation}
where $E_{\mu}$ can be taken~\cite{8,9} to be a shifted Gaussian
weight function with constants $b$ and $c$ as in
$\exp[-b(\mu-c)^2]$, and a similar expression for $E_{\nu}$.

\subsection{The Noncommutative Quantum Cosmology}
\indent

It is well--known~\cite{14} that, in the NC quantum mechanics, the
original phase space and its symplectic structure are modified.
That is, for the NC proposal of quantum cosmology, we assume that
operators (variables) of the FRW minisuperspace obey a kind of
deformed Heisenberg algebra like the ones in the NC quantum
mechanics as~\cite{14}
\begin{equation}\label{A8}
[\hat{x}_i, \hat{x}_j]=i\theta_{ij}\,
,\hspace{.5cm}[\hat{x}_i,\hat{p}_j]=i(\delta_{ij}+\sigma_{ij})
\hspace{.5cm}{\rm
and}\hspace{.5cm}[\hat{p}_i,\hat{p}_j]=i\beta_{ij}\, .
\end{equation}
The notations and definitions are the same as in the NC classical
model. This kind of extended noncommutativity maintains symmetry
between the canonical operators, and yields the usual Heisenberg
algebra in the limit $\theta_{ij}$ and $\beta_{ij}$ $\rightarrow
0$. As usual, this deformation can be redefined in terms of
noncommutativity of minisuperspace functions with the Moyal
product defined in (\ref{A1}). Thus, the corresponding
noncommutative WD equation can be written by replacing the
operator product, in equation (\ref{D1}), with the star product,
namely $\hat{{\cal H}}\ast\Psi=0$.

However, it is possible to reformulate equations in terms of the
commutative operators with the ordinary product of functions if
the new operators are introduced, again as our previous assumption
\begin{equation}\label{A911}
\hat{x}'=
\hat{x}-\frac{\theta}{2}\hat{p}_y\hspace{0.1cm},\hspace{0.5cm}\hat{p}'_x=
\hat{p}_x+2\beta\hat{y} \hspace{0.1cm},\hspace{0.5cm}
\hat{y}'=\hat{y}+\frac{\theta}{2}\hat{p}_x\hspace{0.4cm}
{\rm and} \hspace{0.4cm}\hat{p}'_y= \hat{p}_y-2\beta
\hat{x}\, .
\end{equation}
Clearly, if unprimed operators obey the usual Heisenberg
commutators, then the non--zero primed operators will obey the
deformed Heisenberg commutators (\ref{A8}) in the form
\begin{center}
$[\hat{x}',
\hat{y}']=i\theta,\hspace{.5cm}[\hat{x}',\hat{p}'_x]=i(1+\theta\beta)=[\hat{y}',\hat{p}'_y]\quad
{\rm and }\quad [\hat{p}'_x,\hat{p}'_y]=4i\beta .$
\end{center}
Therefore, operator transformation (\ref{A911}) can be regarded as
a generalization of the usual quantum mechanics to the NC one. On
the other hand, the inverse transformation of (\ref{A911}) are
\begin{center}
$\hat{x}=\eta(\hat{x}'+\frac{\theta}{2}\hat{p}'_y)\hspace{0.1cm},\hspace{0.5cm}\hat{p}_x=\eta(
\hat{p}'_x-2\beta\hat{y}') \hspace{0.1cm},\hspace{0.5cm}
\hat{y}=\eta(\hat{y}'-\frac{\theta}{2}\hat{p}'_x)\hspace{0.4cm}
{\rm and} \hspace{0.4cm}\hat{p}_y= \eta(\hat{p}'_y+2\beta
\hat{x}'),$
\end{center}
where $\eta\equiv 1/(1-\theta\beta)$. Consequently, one can go
from the usual commutators to the deformed ones and vice versa
provided that again $\theta\beta\neq1$. As a result, the original
equation, employing the new operators, reads~\cite{15}
\begin{equation}\label{A92}
\hat{{\cal H}}(\hat{x}_i, \hat{p}_i)\ast\Psi=\hat{{\cal
H}}\left(\hat{x}_i-\frac{1} {2}\theta_{ij}\hat{p}^j,
\hat{p}_i+\frac{1}{2} \beta_{ij}\hat{x}^j\right)\Psi\,=0.
\end{equation}
Hence, the noncommutative WD equation corresponding to the NC
Hamiltonian (\ref{C4}) is
\begin{equation}\label{D9}
\left[\partial^2_a-4ka^2-(1+k\theta^2)(\partial^2_x+\partial^2_y)+
4i(k\theta+\beta)(x\partial_y-y\partial_x)+4(k+\beta^2)(x^2+y^2)\right]\Psi_{\rm
nc}(a,x,y)=0\,,
\end{equation}
which in the polar coordinates (\ref{D2}) yields
\begin{equation}\label{D10}
\left[\partial^2_a-4ka^2-(1+k\theta^2)\left(\partial^2_\rho+\frac{\partial_\rho}{\rho}
+\frac{\partial_\varphi^2}{\rho^2}\right)+4i(k\theta+\beta)\partial_\varphi+4(k+\beta^2)\rho^2\right]\Psi_{\rm
nc}(a,\rho,\varphi)=0.
\end{equation}

When $1+k\theta^2\neq 0$\rlap,\footnote{As will be seen, this
condition should automatically be satisfied when the wave function
is well--behaved.}\
 by using the {\it ansatz} (\ref{D4}), equation (\ref{D10}) is separable to
\begin{equation}\label{D11}
A''+4(\mu-ka^2)A=0,
\end{equation}
and
\begin{equation}\label{D12}
\rho^2B''+\rho
B'+4\left[\frac{\mu+2\nu(k\theta+\beta)}{1+k\theta^2}\rho^2-\frac{k+\beta^2}{1+k\theta^2}\rho^4-
\nu^2\right]B=0,
\end{equation}
where $\mu$ is a constant of separation. The scale factor part of
the wave function, equation (\ref{D11}), is similar to its
commutative analog (\ref{D41}), as expected. The radial part of
the wave function, equation (\ref{D12}), first reduces to its
commutative analog, equation (\ref{D42}), when $\theta=0=\beta$,
as again expected. Secondly, in the case $k=0$ with $\beta=0$,
equation (\ref{D12}) once again reduces to equation (\ref{D42})
even if the NC parameter $\theta$ does exist. However, in $k=0$,
solutions of equation (\ref{D12}) do~not depend on the NC
parameter $\theta$ at all. Namely in a flat FRW universe, the
$\beta$ parameter is the only responsible parameter for the NC
effects. These properties are common with the classical model.

Comparing equation (\ref{D12}) with the commutative analog
(\ref{D42}) shows that the functional form of the radial part of
the wave function (and hence, the whole wave function) is the same
as the commutative ones provided that the coefficients $\mu$ and
$k$ in equation (\ref{D42}) are replaced by
\begin{equation}\label{D121}
\mu\rightarrow\frac{\mu+2\nu(k\theta+\beta)}{1+k\theta^2}\hspace{1cm}\textrm{and}
\hspace{1cm}k\rightarrow\frac{k+\beta^2}{1+k\theta^2}.
\end{equation}
Therefore, for $k=0, 1$ and $k=-1$ with $\theta\neq 1$, the NC
eigenfunctions are in terms of the Whittaker functions similar to
the commutative solutions (\ref{D71}). However, in the case
$k=-1$, there are bounds on the NC parameters. For a better
illustration, let us write (\ref{D121}) as
\begin{center}
$\mu\rightarrow
\sigma\equiv\frac{\mu+2\nu(\beta-\theta)}{1-\theta^2}\hspace{1cm}
\textrm{and} \hspace{1cm}k\rightarrow
K\equiv-\frac{1-\beta^2}{1-\theta^2}.$
\end{center}
Since this modified curvature index, $K$, can have positive,
negative and zero values, the solutions again are similar to the
commutative cases (\ref{C1411}) and (\ref{C14111}), where $K=0$
also requires $\beta=1$. But, for $K\neq 0$, the NC parameters
must satisfy inequalities depending on the sign of $K$. Namely if
$K<0$, then there will be ($\theta$ and $\beta<1$) or ($\theta$
and $\beta>1$), and if $K>0$, then ($\theta>1$ and $\beta<1$) or
($\theta<1$ and $\beta>1$). For instance, by choosing inequality
($\theta$ and $\beta<1$) when $K<0$, the corresponding wave packet
can be written in the form
\begin{equation}\label{D122}
\Psi_{\rm
nc}(a,\rho,\varphi)=\int_{\sigma}^\infty\int_{\sigma}^\infty
E_{\mu}E_{\nu}(\rho \sqrt{a})^{-1}M_{-i\frac{\mu}{2}, \frac{1}{4}}
\left(2i{a}^{2}\right)M_{\frac{\sigma}{2\sqrt{K}},\nu}\left(2\sqrt{K}\hspace{1mm}\rho^2\right)
e^{2i\nu\varphi}d\mu d\nu\,,
\end{equation}
where the lower limits, in contrast to the commutative wave packet
(\ref{D72}), are bounded by $\mu+2\nu(\beta-\theta)\geq 0$.

In the case $1+k\theta^2=0$, namely when $k=-1$ and $\theta=1$
(hence $\beta\neq 1$), with the separation of variables $\Psi_{\rm
nc}=A(a)B(\rho,\varphi)$, equation (\ref{D10}) reduces to two
differential equations. One equation is similar to equation
(\ref{D11}), and the other one is
\begin{equation}\label{D16}
\left[i(1-\beta)\partial_{
\varphi}+(1-\beta^2)\rho^2\right]B(\rho,\varphi)=C\, ,
\end{equation}
where $C$ is a separation constant. As, in general, $C$ has a
non--zero value, equation (\ref{D16}) does~not have a suitable
solution for $\beta=1$ (which itself is~not allowed). For
$\beta\neq1$, one easily shows that its solution can be written as
\begin{center}
$B(\rho,\varphi)=F(\rho)e^{i(1+\beta)\rho^2\varphi}+B_0/\rho^2$\,,
\end{center}
where $F$ is an arbitrary function and $B_0=C/(1-\beta^2)$. As
this function is~not well--behaved when $\rho$ tends to zero, it
is~not allowed.

\section{The Noether Symmetries}
\indent

In this section, we employ the Noether theorem and explore the
effects of noncommutativity on the underlying symmetries in the
commutative frame. For this purpose, following Refs.~\cite{46,17},
one can define the Noether symmetry as a vector field, say $X$, on
the tangent space of the phase space. In our model, it can be, in
general, as
\begin{equation}\label{a}
X=A \frac{\partial}{\partial a}+B \frac{\partial}{\partial x}+C
\frac{\partial}{\partial y}+\frac{d A}{dt}\frac{\partial}{\partial
\dot{a}}+\frac{d B}{dt}\frac{\partial}{\partial \dot{x}}+\frac{d
C}{dt}\frac{\partial}{\partial \dot{y}}\, ,
\end{equation}
such that the Lie derivative of the Lagrangian with respect to
this vector field vanishes, i.e.
\begin{equation}\label{b}
L_X{\cal L}=0.
\end{equation}
For simplicity, we assume that unknown functions $A$, $B$ and $C$
to be linear in terms of the configuration variables $a$, $x$ and
$y$. The time derivative $d/dt$ represents the Lie derivative
along the dynamical vector fields, which in our model is
$d/dt=\dot{a}\partial/\partial a+\dot{x}\partial/\partial
x+\dot{y}\partial/\partial y$.

Now, in order to obtain constants of motion, let us rewrite
equation (\ref{b}) as
\begin{equation}\label{d}
\left(A\frac{\partial {\cal L}}{\partial
a}+\frac{dA}{dt}\frac{\partial {\cal L}}{\partial
\dot{a}}\right)+\left(B \frac{\partial {\cal L}}{\partial
x}+\frac{dB}{dt}\frac{\partial {\cal L}}{\partial
\dot{x}}\right)+\left(C \frac{\partial {\cal L}}{\partial
y}+\frac{dC}{dt}\frac{\partial {\cal L}}{\partial
\dot{y}}\right)=0.
\end{equation}
By employing the Lagrange equation, $\partial\cal L/$$\partial q$=
$dp_q/dt$, it reads
\begin{equation}\label{e}
\left(A
\frac{dp_a}{dt}+\frac{dA}{dt}p_a\right)+\left(B
\frac{dp_x}{dt}+\frac{dB}{dt}p_x\right)+\left(C
\frac{dp_y}{dt}+\frac{dC}{dt}p_y\right)=0,
\end{equation}
which yields
\begin{equation}\label{f}
\frac{d}{dt}\left(A p_a+B p_x+C p_y\right)=0.
\end{equation}
Therefore, the constants of motion are
\begin{equation}\label{g}
Q\equiv Ap_a+Bp_x+Cp_y,
\end{equation}
for different unknown functions of $A$, $B$ and $C$. To obtain
these functions, one can employ equation (\ref{d}) or equation
(\ref{e}) which is more suitable in the Hamiltonian formalism. To
manage this, one can write equation (\ref{e}) in terms of the
Poisson bracket, $\{\hspace{2mm},H\}=d/dt$, as
\begin{eqnarray}\label{h}
A\{p_a,{\cal H}\}+B\{p_x,{\cal H}\}+C\{p_y,{\cal
H}\}+\left[\frac{\partial A}{\partial a}\{a,{\cal
H}\}+\frac{\partial A}{\partial x}\{x,{\cal H}\}+\frac{\partial
A}{\partial y}
\{y,{\cal H}\}\right]p_a\nonumber\\
+\left[\frac{\partial B}{\partial a}\{a,{\cal H}\}+\frac{\partial
B}{\partial x}\{x,{\cal H}\}+\frac{\partial B}{\partial y}
\{y,{\cal H}\}\right]p_x+ \left[\frac{\partial C}{\partial
a}\{a,{\cal H}\}+\frac{\partial C}{\partial x}\{x,{\cal
H}\}+\frac{\partial C}{\partial y} \{y,{\cal H}\}\right]p_y=0.
\end{eqnarray}
This equation, in general, gives quadratic polynomials in terms of
the momenta with coefficients being partial derivatives of $A$,
$B$ and $C$ with respect to the configuration variables. Hence,
the expression identically is equal to zero if and only if these
coefficients vanish, which lead to a system of partial
differential equations for $A$, $B$ and $C$.

In the following subsections, we obtain such symmetries for the
model in the commutative and NC cases.

\subsection{Symmetries in The Commutative Frame}
\indent

In this case the Hamiltonian is given by relation (\ref{B5,2});
hence by substituting the corresponding Poisson brackets into
equation (\ref{h}), one gets
\begin{eqnarray}\label{i}
2k(a A-x B-y C)\nonumber\\
+\frac{1}{2}\left(-\frac{\partial A}{\partial
a}p_a^2+\frac{\partial B}{\partial
x}p_x^2+ \frac{\partial C}{\partial y}p_y^2\right)\nonumber \\
+\frac{1}{2}\left(\frac{\partial A}{\partial x}-\frac{\partial
B}{\partial a}\right)p_ap_x+ \frac{1}{2}\left(\frac{\partial
A}{\partial
y}-\frac{\partial C}{\partial a}\right)p_ap_y
+\frac{1}{2}\left(\frac{\partial B}{\partial y}+\frac{\partial
C}{\partial x}\right)p_x p_y=0.
\end{eqnarray}
Let us first treat spatially non--flat geometry, $k\neq 0$, for
which equation (\ref{i}) leads to
\begin{eqnarray}\label{j}
a A-x B-y C=0,\nonumber\\
\frac{\partial A}{\partial a}=
\frac{\partial B}{\partial x}=\frac{\partial C}{\partial y}=0,\nonumber\\
\left(\frac{\partial A}{\partial x}-\frac{\partial
B}{\partial a}\right)=\left(\frac{\partial
A}{\partial
y}-\frac{\partial C}{\partial a}\right)=
\left(\frac{\partial B}{\partial y}+\frac{\partial
C}{\partial x}\right)=0.
\end{eqnarray}
A general solution of constraints (\ref{j}) is
\begin{equation}\label{l}
A=c_1x+c_2y\, ,\qquad B=c_1a-c_3y\qquad {\rm and} \qquad
C=c_2a+c_3x\, ,
\end{equation}
where $c_i$'s are three constants of integration. For an easier
representation of the commutation relations between symmetric
vectors, let us change $x\rightarrow -x$ and $y\rightarrow -y$ in
solution (\ref{l}). Thus by (\ref{g}), three independent constants
of motion are
\begin{equation}\label{m}
Q_1=ap_x-x p_a\, ,\qquad Q_2=ap_y-y p_a\qquad {\rm and} \qquad
Q_3=y p_x-x p_y\, ,
\end{equation}
which are the well--known angular momenta about the configuration
variables. The corresponding symmetric vectors are
\begin{eqnarray}\label{ae}
X_1\!\!&=&\!\! a\frac{\partial}{\partial
x}-x\frac{\partial}{\partial a}+\dot{a}\frac{\partial}{\partial
\dot{x}}-\dot{x}\frac{\partial}{\partial \dot{a}}\, ,
\nonumber \\
X_2\!\!&=&\!\! a\frac{\partial}{\partial
y}-y\frac{\partial}{\partial a}+\dot{a}\frac{\partial}{\partial
\dot{y}}-\dot{y}\frac{\partial}{\partial \dot{a}}\nonumber\, ,
\\
X_3\!\!&=&\!\! y\frac{\partial}{\partial
x}-x\frac{\partial}{\partial y}+\dot{y}\frac{\partial}{\partial
\dot{x}}-\dot{x}\frac{\partial}{\partial \dot{y}}\, ,
\vspace{.5cm} \nonumber\\
\end{eqnarray}
which satisfy the Lie algebra
$[X_i,X_j]=\varepsilon_{ij}{}^{k}X_k$\,, where $\varepsilon_{ijk}$
is the Levi--Civita tensor.

For $k=0$, it is clear from equation (\ref{i}) that the symmetries
can be obtained from the last two constraints of (\ref{j}). Thus,
the corresponding solution is
\begin{equation}\label{ac}
A=d_1x+d_2y+d_3\, ,\qquad B=d_1a+d_4y+d_5\qquad {\rm and}\qquad
C=d_2a-d_4x+d_6\, ,
\end{equation}
where $d_i$'s are six constants of integration. Therefore, the six
independent constants of motion are
\begin{equation}\label{ad}
\begin{array}{c}
Q_1=p_a\,,\hspace{2.2cm}Q_2=p_x\hspace{1mm},\hspace{2.8cm}Q_3=p_y\hspace{1mm},\vspace{.5cm} \\
Q_4=y p_x-x p_y\hspace{1mm},\hspace{.8cm}Q_5=y
p_a+ap_y\hspace{.8cm}{\rm and} \hspace{.8cm}Q_6=x
p_a+ap_x\hspace{1mm}.
\end{array}
\end{equation}
Incidentally, in this case, a glance at Hamiltonian (\ref{B5,2})
shows that all configuration variables are cyclic and consequently
their corresponding momenta are constants of motion, i.e. $Q_1$ to
$Q_3$. In addition, Hamiltonian (\ref{B5,2}) for $k=0$ (actually
the Lagrangian $\dot{x}^2+\dot{y}^2-\dot{a}^2$) is invariant under
rotation about the $a$--axis. Thus, the angular momentum about
this axis is conserved, i.e. $Q_4$.

The corresponding symmetric vectors, in the flat geometry, are
\begin{eqnarray}
X_1&=&\frac{\partial}{\partial
a}\hspace{1mm},\hspace{1mm}X_2=\frac{\partial}{\partial x}\hspace{1mm},
\hspace{1mm}X_3=\frac{\partial}{\partial y}\hspace{1mm},\nonumber \\
X_4&=&y\frac{\partial}{\partial x}-x\frac{\partial}{\partial
y}+\dot{y}\frac{\partial}{\partial
\dot{x}}-\dot{x}\frac{\partial}{\partial \dot{y}}\hspace{1mm},\nonumber\\
X_5&=&y\frac{\partial}{\partial a}+a\frac{\partial}{\partial
y}+\dot{y}\frac{\partial}{\partial
\dot{a}}+\dot{a}\frac{\partial}{\partial \dot{y}}\hspace{1mm},
\nonumber \\
X_6&=&x\frac{\partial}{\partial a}+a\frac{\partial}{\partial
x}+\dot{x}\frac{\partial}{\partial
\dot{a}}+\dot{a}\frac{\partial}{\partial \dot{x}}\hspace{1mm},\nonumber
\end{eqnarray}
which satisfy
\begin{eqnarray}\label{af1}
\left[X_1,X_2\right]&=&0,\hspace{1.5cm}[X_1,X_3]=0,\hspace{1.3cm}[X_1,X_4]=0,\nonumber\\
\left[X_1,X_5\right]&=&X_3,\hspace{1.2cm}[X_1,X_6]=X_2,\hspace{1cm}[X_2,X_3]=0,\nonumber \\
\left[X_2,X_4\right]&=&-X_3,\hspace{.9cm}[X_2,X_5]=0,\hspace{1.3cm}[X_2,X_6]=X_1,\\
\left[X_3,X_4\right]&=&X_2,\hspace{1.2cm}[X_3,X_5]=X_1,\hspace{1cm}[X_3,X_6]=0,\nonumber \\
\left[X_4,X_5\right]&=&-X_6,\hspace{.9cm}[X_4,X_6]=X_5,
\hspace{1cm}[X_5,X_6]=X_4.\nonumber
\end{eqnarray}

\subsection{Symmetries in The Noncommutative Frame}
\indent

Now, let us find out which of the above symmetries survive in the
NC case. Here, the Hamiltonian is given by relation (\ref{C4}),
and the required Poisson brackets are given by equations
(\ref{C5}) and (\ref{C9}). Substituting the corresponding Poisson
brackets into equation (\ref{h}) gives
\begin{eqnarray}\label{i1}
2kaA-2(k+\beta^2)(x B+y C)\nonumber \\
 +\frac{1}{2}\left[-\frac{\partial A}{\partial
a}p_a^2+\frac{1}{2}(1+k\theta^2)\left(\frac{\partial B}
{\partial x}p^2_x+\frac{\partial C}{\partial y}p^2_y\right)\right]\nonumber \\
 +\frac{1}{2}\left[(1+k\theta^2)\frac{\partial A}{\partial
x}-\frac{\partial B}{\partial
a}\right]p_ap_x+\frac{1}{2}\left[(1+k\theta^2)\frac{\partial
A}{\partial y}-\frac{\partial C}{\partial
a}\right]p_ap_y+\frac{1}{2}(1+k\theta^2)\left(\frac{\partial
B}{\partial y}+ \frac{\partial C}{\partial x}\right)p_x p_y\nonumber\\
 +(k\theta+\beta)\left[\left(y\frac{\partial A}{\partial
x}-x\frac{\partial A}{\partial y}\right)p_a+\left(y\frac{\partial B}{\partial
x}-x\frac{\partial B}{\partial y}-C\right)p_x+\left(y\frac{\partial C}{\partial
x}-x\frac{\partial C}{\partial y}+B\right)p_y\right]=0.
\end{eqnarray}
Obviously, equation (\ref{i1}) yields extra restrictions on $A$,
$B$ and $C$ with respect to the commutative case (\ref{i}).

When $k\theta+\beta\neq 0$\rlap,\footnote{The equality
$k\theta+\beta=0$ is possible when ($k=-1$ and $\theta=\beta$) and
or ($k=0=\beta$ and any value of $\theta$).}\
 by putting each coefficient in equation (\ref{i1}) equal to zero, one gets
\begin{eqnarray}\label{p1}
kaA-(k+\beta^2)(x B+y C)=0,\nonumber\\
 \frac{\partial A}{\partial a}=\frac{\partial B}{\partial x}=\frac{\partial C}{\partial y}=0,\nonumber\\
 (1+k\theta^2)\frac{\partial A}{\partial x}-\frac{\partial
B}{\partial a}=(1+k\theta^2)\frac{\partial A}{\partial
y}-\frac{\partial C}{\partial a}=\frac{\partial
B}{\partial y}+ \frac{\partial C}{\partial x}=0,\nonumber\\
 y\frac{\partial A}{\partial
x}-x\frac{\partial A}{\partial y}=y\frac{\partial B}{\partial
x}-x\frac{\partial B}{\partial y}-C=y\frac{\partial C}{\partial
x}-x\frac{\partial C}{\partial y}+B=0.
\end{eqnarray}
Hence, one obtains the solution
\begin{equation}\label{q}
B=h_0\, y\,, \hspace{2cm}C=-h_0\,x
\end{equation}
and
\begin{eqnarray}\label{qr}
A=\left\{
\begin{array}{lll}
A_0 \hspace{1cm} {\rm for}\hspace{0.6cm}k=0
\\
0\hspace{1.3cm}{\rm for}\hspace{0.6cm} k\neq0\, ,
\end{array}
\right.
\end{eqnarray}
where $h_0$ and $A_0$ are integration constants. Thus, when $k\neq
0$, the only constant of motion is
\begin{equation}\label{r}
Q_{\rm nc}=y p_x-x p_y\hspace{.5mm},
\end{equation}
with the symmetric vector
\begin{equation}\label{s}
X_{\rm nc}=y\frac{\partial}{\partial x}-x\frac{\partial}{\partial
y}+\dot{y}\frac{\partial}{\partial
\dot{x}}-\dot{x}\frac{\partial}{\partial \dot{y}}\, ,
\end{equation}
which is an especial case of commutative solution (\ref{l}) with
initial conditions $c_1=0=c_2$. For $k=0$ (with $\beta\neq 0$), in
addition to $Q_{\rm nc}$, we have an another constant of motion,
namely $Q_a=p_a$, which corresponds to the symmetric vector
$X_a=\partial/\partial a$ with the trivial Lie algebra $[X_{\rm
nc},X_a]=0$. This case also restricts commutative solution
(\ref{ac}) with initial conditions $d_1=d_2=d_5=d_6=0$. Therefore,
in the $k\theta+\beta\neq 0$ case, the noncommutativity reduces
the number of symmetries to one for $k\neq 0$ and two for $k=0$.

When $k\theta+\beta=0$, the last row of constraints in (\ref{p1})
is omitted. In this case, one choice is $k=-1$ and $\theta=\beta$,
where a non--trivial (i.e. $\theta=\beta\neq 0$) linear solution
exists if and only if $\theta=\sqrt{2}=\beta$. Hence, by the sign
change of functions, i.e. $B\rightarrow -B$ and $C\rightarrow -C$
(or equivalently $A\rightarrow -A$), the solution is the same as
the commutative solution (\ref{l}), which again yields the
constants of motion and symmetric vectors (\ref{m}) and
(\ref{ae}). However for $k=-1$ and a special value of
$\theta=1=\beta$, it gives one solution (i.e. (\ref{r})), but this
case is~not allowed due to the inversion condition
$\theta\beta\neq 1$. For another choice $k=0=\beta$ with any value
of $\theta$, it gives the commutative solution (\ref{ac}). That
is, even though the noncommutativity is still present, the number
of symmetries and constants of motion do~not change with respect
to the corresponding commutative case.

Note that, in all cases, irrespective of whether the
noncommutativity exists or not, and for any value of the curvature
index, the angular momentum about the $a$--axis is conserved, as
expected. Besides, in the above considerations, we have~not, in
general, specified numerical values of the NC parameters.

\section{Conclusions}
\indent

We have carried out an investigation for the role of NCG in
cosmological scenarios, based on a four--dimensional
free--potential (multi)scalar--tensor action of gravity, by
introducing a NC deformation in the minisuperspace variables. The
phase space is generated by two non--interacting conformal scalar
fields plus the scale factor with their canonical conjugate
momenta. The scalar fields are non--minimally coupled to geometry
whose background is the FRW metric, where the conformal time gauge
evolutions have been studied. The noncommutativity has been
introduced only between the scalar fields and between their
canonical conjugate momenta via two NC parameters $\theta$ and
$\beta$, respectively. The investigation has been carried out for
this toy model by means of a comparative mathematical analysis of
the time evolution of the dynamical variables in the classical
level and of the wave function of the universe in the quantum
perspective, both in the commutative and NC frames. We have paid
more attention to the outcome of results and have looked for the
relations, including ranges and values, among the NC parameters
for which particular or allowed solutions exist.

In general, we have shown by this toy model that the purposed
noncommutativity has important implications in the evolution of
the universe, however, does~not affect the time dependence of the
scale factor, i.e. its solution is the same as the commutative
case, as expected. Also, we have found that one of the
particularity of the NC parameter in the momenta sector, i.e.
$\beta$, is in the spatially flat FRW universe, where it is the
only responsible parameter for the NC effects in the classical and
quantum frames.

In the classical model, exact solutions have been obtained. One of
the important aspects of the NC solutions is that they can be
regulated with both NC parameters. For example, these parameters
can be employed to adjust the time dependence of solutions with
the experimental or observational data. A distinguished feature of
the noncommutativity effects, which we call a cosmical oscillator,
is that the scalar fields behave similar to (or can be simulated
with) the three most important harmonic oscillators depending on
three values of the spatial curvature. These are the free, forced
and damped harmonic oscillators corresponding to the flat, closed
and open universes, respectively. In the flat universes, the time
dependence of solutions are modified from linear in the
commutative case to periodic in the NC frame. In the closed
universes, if the ration of frequencies of the scalar fields is a
rational fraction, then solutions will be periodic. This condition
restricts the NC parameters. When this ratio is~not a rational
fraction, the solutions are non--periodic but their behaviors
still depend on the values of the NC parameters. A plot with
numerous (or a few) relative extremum in a given time interval
indicates that the separation between the high and low points
increases when the NC parameters tend to smaller values, and this
property is intensified for values less than unity. From this
point of view, the NC solutions have particular preference with
respect to the corresponding commutative ones. Furthermore, in the
$k=1$ case, the solutions are symmetric with respect to the NC
parameters, and the results do~not vary by replacement of their
roles. Also, when the NC parameters are small, the cosmical
oscillators have analogous effects with the familiar beating
effects in the sound phenomena. For a better view on this
situation, an example has been illustrated in the text. In the
open universes, there are upper and lower bounds on the NC
parameters. The quick and late damping of this case can be
adjusted by the NC parameters. For example, for $\theta\rightarrow
1^-$\ and $\beta\rightarrow 0$, the scalar fields have maximum
number of oscillations before the complete damping occurs.

We have also shown that the existence of a non--zero constant
value of the potential function (i.e. the cosmological constant)
does~not change the time evolutions of the scalar fields, but it
modifies the time dependence of the scale factor in a same manner
for both the commutative and NC frames. Indeed, we have obtained
that in all allowed conditions the scale factor behaves as a
sinusoidal function.

In the quantum model, the exact solutions for the wave functions
of the universe have also been obtained. The scale factor part of
the wave function is similar to its commutative analog, as
expected. One still expects that when the noncommutativity effects
are turned on in the quantum scenario, they should introduce
significant modifications in the scalar fields. However, an
interesting feature of the well--behaved solutions is that the
functional form of the radial part of the NC wave function is the
same as the commutative ones within the given replacements of
constants caused by the NC parameters, although these replacements
in turn may cause drastic effects. For example, the curvature
index is modified, and in open universes, the allowed NC wave
functions impose bounds on the NC parameters.

Finally, we have employed the Noether theorem and have explored
the effects of noncommutativity on the underlying symmetries in
the commutative frame. We have shown that for spatially flat
universes, there are six Noether symmetries, and, in general, only
two of them are retained in the NC case. In the special case
$k=0=\beta$, all symmetries survive regardless of the existence of
$\theta$ parameter. For non--flat universes, there are three
Noether symmetries in the commutative case, one of which is
retained in the NC frame. However, in open universes, when the NC
parameters have the values $\theta=\sqrt{2}=\beta$, the number and
general form of symmetries do~not change with respect to the
commutative frame. The only difference is related to the sign of
symmetries. Conservation of the angular momentum about the scale
factor axis is a common face between the commutative and NC cases,
as expected in the purposed noncommutativity.

We should emphasis that the scalar fields solutions are given
after rescalings of the original fields, relations
(\ref{rescalingfields}), and in the conformal time. Consequently,
the functional forms of the solutions are~not as simple when
translated to the original fields. However, any detection should
be performed in a reference frame, and there are debates on the
physical frame in the cosmological contexts. Besides, the model is
to be taken as a toy model only that still provides a valuable
contribution to the assessment of the implications of the NC
geometrical deformations of the phase space upon the dynamics of
the cosmological model envisaged. However, a more realistic NC
cosmological model may be achieved when one also involves the
noncommutativity of the scalar fields with the scale factor, where
the value of its time derivative in the form of the Hubble
parameter can be determined from observations.

\setcounter{equation}{0}
\renewcommand{\theequation}{A.\arabic{equation}}
\section*{Appendix}
\indent

We first obtain a general relation among the NC parameters when
the solutions (\ref{C16}) are periodic. Then, we treat a specific
periodic solution as an example.

As mentioned in the text when $k=1$, solutions (\ref{C16}) play as
the forced oscillators, and the existence of periodic solutions is
provided when the ratio $\omega_2/\omega_1$ is a rational
fraction. In this case, the $\Delta$ parameter, by its definition,
is $\Delta =(\theta+\beta)^2+(\theta\beta-1)^2$; hence by the
transformation reversibility constraint $\theta\beta\neq 1$, one
always has $\sqrt{\Delta}>\theta+\beta$. Therefore, by definitions
(\ref{C161}), we can take $\omega_1=cn_1$ and $\omega_2=-cn_2$,
where $n_1$ and $n_2$ are two positive integers with $n_2>n_1$,
and $c$ is an arbitrary non--zero positive constant. Hence, again,
definitions (\ref{C161}) give
\begin{equation}\label{CA}
\theta+\beta=\frac{c}{2}(n_2-n_1)\hspace{1cm} {\rm and}
\hspace{1cm} \theta\beta=1\pm c\sqrt{n_1n_2}\geq 0\, .
\end{equation}
These relations impose a firm restriction on the values of the NC
parameters (or equivalently on allowed positive integers). Indeed,
the condition for periodic scalar fields is that, the values of
the NC parameters must be such that there exist two positive
integers satisfying relations (\ref{CA}). As $\theta$ and $\beta$
are real roots of the quadratic equation
$\textsl{X}^2-(\theta+\beta) \textsl{X}+\theta\beta=0$, relations
(\ref{CA}) lead to
\begin{equation}\label{CAA}
(n_2-n_1)^2\geq(16/c^2)(1\pm c\,\sqrt{n_1n_2}\,).
\end{equation}
The above result implies that if one finds a pair of integers
satisfying inequality (\ref{CAA}), then the values of the NC
parameters will be calculated from equations (\ref{CA}).

Let us treat a specific example for the above considerations.
Consider the equality in relation (\ref{CAA}), which yields
$\theta=\beta$, which also by the constraint $\theta\beta\neq 1$
gives $\theta\neq 1$ (note that we have assumed $\theta\geq 0$).
Also, by definitions (\ref{C161}), we have $\omega_1=(1-\theta)^2$
and $\omega_2=-(1+\theta)^2$. Thus, by imposing the periodic
condition as
\begin{equation}\label{CAAA}
-\frac{\omega_2}{\omega_1}=\left(\frac{1+\theta}{1-\theta}\right)^2={\rm
rational\hspace{2mm} fraction}\equiv m^2>1,
\end{equation}
it leads to
\begin{equation}\label{CAAAA}
\theta=\frac{m+1}{m-1}>1\qquad\quad {\rm or}\qquad\quad
\theta=\frac{m-1}{m+1}<1\, ,
\end{equation}
where $m$ is a real number greater than
one\rlap.\footnote{Clearly, the value $m=1$ gives the commutative
case.}\
 In the limit $m\rightarrow 1$, one solution of (\ref{CAAAA})
yields $\theta\rightarrow 0$, as expected; however, the other one
gives infinity that is~not physically accepted. Also, in the limit
$m\rightarrow \infty$, we have $\theta\rightarrow 1$, which, in
general, may be viewed as a transition from the forced harmonic
oscillator in the $k=1$ geometry to the simple one in the $k=0$
case. In another words, when the value of $m$ slowly gets bigger
and bigger, then the driven force is gradually removed away from
the cosmical oscillators. The time evolution of scalar field
$x(t)$, equation (\ref{C16}), is depicted in Fig.~$2$ (left) for
the numerical value $m=3$ corresponding to $\theta=2$, with
superposition constants $A=2$ and $B=1$. The period of solutions
is the least common multiple of distinct periods, namely
$2\pi/(1-\theta)^2$ and \hspace{1mm} $2\pi/(1+\theta)^2$.

\section*{Acknowledgement}
\indent

We thank the Research Office of the Shahid Beheshti University for
financial support. This work is a part of B.M.'s Ph.D. thesis
under supervision of M.F..
%%%%%%%%%%%%%%%%%%%%%%%%%%%%%%%%%%%%%%%%%%%%%%%%%%%%%%%%%%%%%%%%%%%%

%
\end{document}